\magnification=1200
\def\bfb{{\bf b}}
\def\bfv{{\bf v}} 
\def\bfbb{{\bar{\bf B}}}
\def\bfbv{{\bar{\bf V}}}

\def\.{\mathaccent 95}
\def\a{\alpha}
\def\be{\beta}

\def\de{\delta}
\def\ep{\epsilon}

\def\Ga{\Gamma}

\def\frac#1#2{{\textstyle{{#1}\over {#2}}}}

\def\lsim{\mathrel{\rlap{\lower4pt\hbox{\hskip1pt$\sim$}}
    \raise1pt\hbox{$<$}}}
\def\gsim{\mathrel{\rlap{\lower4pt\hbox{\hskip1pt$\sim$}}
    \raise1pt\hbox{$>$}}}
\def\sqr#1#2{{\vcenter{\vbox{\hrule height.#2pt
         \hbox{\vrule width.#2pt height#1pt \kern#1pt
         \vrule width.#2pt}
         \hrule height.#2pt}}}}

\newbox\grsign \setbox\grsign=\hbox{$>$} \newdimen\grdimen \grdimen=\ht\grsign
\newbox\simlessbox \newbox\simgreatbox
\setbox\simgreatbox=\hbox{\raise.5ex\hbox{$>$}\llap
     {\lower.5ex\hbox{$\sim$}}}\ht1=\grdimen\dp1=0pt
\setbox\simlessbox=\hbox{\raise.5ex\hbox{$<$}\llap
     {\lower.5ex\hbox{$\sim$}}}\ht2=\grdimen\dp2=0pt

%
%

\def\ref#1  {\noindent \hangindent=24.0pt \hangafter=1 {#1} \par}
\def\doublespace {\smallskipamount=6pt plus2pt minus2pt
                  \medskipamount=12pt plus4pt minus4pt
                  \bigskipamount=24pt plus8pt minus8pt
                  \normalbaselineskip=24pt plus0pt minus0pt
                  \normallineskip=2pt
                  \normallineskiplimit=0pt
                  \jot=6pt
                  {\def\smallskip {\vskip\smallskipamount}}
                  {\def\medskip   {\vskip\medskipamount}}
                  {\def\bigskip   {\vskip\bigskipamount}}
                  {\setbox\strutbox=\hbox{\vrule 
                    height17.0pt depth7.0pt width 0pt}}
                  \parskip 12.0pt
                  \normalbaselines}
\def\ts{\times}
\def\lb{\langle}
\def\rb{\rangle}
\def\curl{\nabla {\ts}}

\def\intt{\int^t_{0}dt'}

\doublespace

\centerline {\bf OVERCOMING THE BACK REACTION}
\centerline {\bf ON TURBULENT MOTIONS IN THE PRESENCE OF MAGNETIC FIELDS}
\medskip
\centerline{\bf Eric G. Blackman}
\smallskip
\centerline{Institute of Astronomy, Madingley Road, Cambridge, CB3 OHA,
England}
\smallskip
\medskip

\centerline{(accepted for publication in {\it Physical Review Letters})}

\centerline{\bf ABSTRACT}

Standard magnetohydrodynamic theories, such as
the mean field dynamo theory, have been 
criticized when the back reaction of the magnetic 
field on turbulent motions is neglected.  For the dynamo, 
this back reaction has been argued to suppress the turbulent motions
required for optimal mean field production.  Here it is suggested 
that if the magnetic field 
is spatially intermittent, for example residing in flux tubes, the 
back reaction on turbulent flows may be significantly reduced.

\medskip

\vfill
\eject

{\it 1. Introduction-}
The Mean Field Dynamo (MFD), an elegant 
theoretical mechanism that allows a large scale magnetic field to 
grow exponentially at the expense of shear and small scale turbulent energy, 
has been studied to explain the origin of 
magnetic fields in astrophysical bodies such as planets, stars, 
and galaxies [1,2,3,4].  The difficulties with
dynamo physics highlight fundamental
issues of magneto-hydrodynamics that are still not well understood.
In particular, the effect of magnetic fields on turbulent flows.
 
In principle, the  MFD growth of a 
large scale magnetic field in a differentially 
rotating system occurs as turbulent 
motions induce formation of magnetic loops from a  seed 
field [1,2,3].  As a result of the coriolis and buoyancy forces, the 
turbulence is cyclonic; the loops in both hemispheres twist in the 
same direction, creating a large scale loop of poloidal field.  The 
turbulent diffusion of the outer portions of these mean field loops ensures 
that the net flux of mean field lying in the region of interest is 
non-vanishing [1].  Differential rotation shears the 
large scale poloidal field, generating a large scale toroidal field.  
The new field then incurs the same loop forming 
process, providing the feedback for exponential growth.
The strength of the field is limited by the available turbulent energy.
Numerical simulations of ``kinematic'' dynamos in which the
back reaction of the field growth on the turbulent eddies is neglected, 
can produce magnetic topologies consistent with observations of stellar and galactic 
magnetic fields [5,6].

In reality, the small scale, root mean squared (rms)
 field energy density grows much more rapidly than the mean field [1], 
violating the kinematic approximation.  
Though, for example, observations of the solar photosphere and the 
dispersed heavy element
distribution in the Galaxy  
indicate the presence of reasonably uninhibited turbulent motions
in amidst equipartition magnetic fields [1], theoretical 
studies [7,8]  suggest
that the back reaction should suppress these motions and thereby 
inhibit the dynamo.  
As first discussed by Piddington [9], the argument proceeds as follows:  
As an eddy tries to displace a parcel of fluid threaded by a field loop
in equipartition with the turbulent energy density, 
the loop acts as a rubber band, restricting further motion, 
and recoiling the parcel to its origin.  Statistically,  transport of 
material is significantly reduced.  Simulations support this 
intuition by finding that an increasingly 
large fraction of plasma motions are locked into oscillating modes
rather than zero frequency (diffusive) modes as the magnitude of 
the initial ordered field is increased [8].  

Here it is suggested that turbulent 
motion may survive the back reaction 
if the magnetic field is concentrated in flux tubes.
Because the  tubes' Alfv\'en speeds can be larger than the eddy 
velocities, reconnection between tubes can be rapid.  
This would  reduce
the back reaction force on the turbulent velocities.  
The diffusion and helicity coefficients of the MFD equation would be 
reduced from the kinematic theory only by the fraction of a typical 
tube thickness which does not reconnect in an eddy turnover time.  
(In the special case for which the magnetic field  is 
totally composed of topologically unlinked 
loops or cells, interchange motions between
the cells would not be restricted by magnetic forces and reconnection
would not be required for diffusive motions.)


First, the derivation of the dynamo helicity and diffusion 
coefficients is outlined and the specific
role of the kinematic approximation is highlighted.
The stages of dynamo growth are discussed and 
the formation and role of flux tubes is then addressed.
An estimate of the dynamo coefficients is given based on the physical
ideas presented.   Finally, a similar role of intermittency for star 
forming regions is mentioned.




{\it 2. Dynamo Coefficients and Interpretation of Approximations-}
The magnetic field and velocity are taken to be 
${\bf B}={\bf b}+{\bar{\bf B}}$ and ${\bf V}={\bf v}+{\bar {\bf
V}}$ respectively, where ${\bf b}$ and ${\bf v}$ are fluctuating
quantities with zero mean and ${\bar {\bf B}}$ and ${\bar {\bf V}}$
are mean quantities.
The Reynolds relations [11] are also required.  
These are  $\partial_{t,{\bf x}}\lb {K_i H_j} \rb=
\lb\partial_{t,{\bf x}}({K_i H_j})\rb$, and $\lb{\bar {K_i}} {h_j}\rb=0$, 
where $K_i={\bar K}_i+k_i$ and $H_j={\bar H}_j=h_j$ are components of 
vector functions of position $\bf x$ and time $t$, the brackets indicate 
the mean value, and $\partial_{t,{\bf x}}$ is the derivative 
with respect to $\bf x$ or $t$. For ensemble averages, 
these relations hold when the turbulence is correlated on time scales short
relative to the variation time scales of the mean quantities. 
For the spatial average, defined by $\lb {\bf K}({\bf x},t)\rb = |{\bf \zeta}|^{-3}\int^{{\bf x}+
{\bf \zeta}}_{{\bf x}-{\bf \zeta}}{\bf K}({\bf s},{\rm t}){\rm d^3}{\bf s}$,
these relations hold when the average is taken over
a large enough scale, that is 
when $l<<|{\bf \zeta}|<<L$, where
$L\sim {\bar B}/\nabla {\bar B}$ is the scale of the mean
field variation, and $l\sim b/\nabla b\sim v/\nabla v$.



The induction equation derived from the non-relativistic Maxwell
equations is [1] 
$$\partial({\bar {\bf B}}+{\bf b})/\partial t=\nabla {\ts}[({\bar {\bf V}}+{\bf v}){\ts}({\bar {\bf B}}+{\bf b})]+\nu_M\nabla^2({\bar {\bf B}}+{\bf b}),\eqno(1)$$
where $\nu_M$ is the magnetic viscosity.  
Astrophysical magnetic 
Reynolds numbers are large and the last 
term in (1) is unimportant on the energy
containing eddy scales for destroying
magnetic field energy  (but it does provide dissipation on the smaller scales 
allowing a turbulent cascade, and
is important  locally, at the intersections between thin flux tubes.)
 Taking the average of $(1)$, ignoring the last term, gives the MFD equation 
$$\partial{\bar {\bf B}}/\partial t=\nabla {\ts}[{\bar{\bf V}}\ts{\bar
{\bf B}}+\lb{{\bf v}}{\ts}{\bf b}\rb].\eqno(2)$$
The turbulent electromotive force (TEMF) given by
 $\lb\bfv\ts\bfb\rb$, is written [2] 
$$ \lb\bfv\ts\bfb\rb={\tilde \a}_{ij}{\bfbb_j}+{\tilde \beta}_{ijk}\nabla_j\bfbb_k,\eqno(3)$$
where ${\tilde \a}_{ij}$ is the helicity dynamo coefficient, and ${\tilde \beta}_{ijk}$ is the turbulent diffusion dynamo coefficient.  Under the 
assumption of isotropic turbulence
${\tilde \a}_{ij}=\de_{ij}\tilde \a$ and ${\tilde \beta}_{ijk}=
\ep_{ijk}{\tilde \beta}$.

Working in a frame for which ${\bar {\bf V}}=0$, 
subtracting $(2)$ from $(1)$ gives 
$$\partial{\bf b}/\partial t=\nabla {\ts}({\bf v}{\ts}{\bar {\bf
B}})+
\nabla{\ts}[{\bf v}{\ts}{\bf b}-\lb{\bf v}{\ts}{\bf
b}\rb]+\bfb\cdot\nabla{\bfbv}.
\eqno(4)$$
Usually, for simplicity, the 
3rd and 4th terms of (4) are neglected straight away, which is
a procedure called the ``first order smoothing (FOS) [1,2]'' approximation.
Instead, I will keep these terms and then show what this approximation means.
Plugging  (4) into (3),
the dynamo coefficients become expansions
of  time-ordered exponential series [13] in powers of
$\tau_c/\tau_{ed}$,
where $\tau_c$ is the turbulence correlation time and $\tau_{ed}\sim l/v$
is the eddy turnover time.
Under the assumption of isotropy, 
taking the first terms in the series give the standard forms [1,2]
$${\tilde \a}= (-1/3)\lb \bfv(t)\cdot\intt \curl \bfv(t')\rb,\eqno(5)$$
$${\tilde \beta}= (1/3)\lb\bfv(t)\cdot\intt\bfv(t')\rb.\eqno(6)$$
Non-vanishing $\tilde \a$ means non-vanishing helicity, essential for mean field 
growth as described above.
That ${\tilde \beta}$ acts as a diffusion coefficient for the mean field is
evident from plugging (6) into the TEMF in (2).  For spatially homogeneous
turbulence, ${\tilde \beta}$  becomes the coefficient of the diffusion term
on the right side.  

Note that dropping the higher order terms that led to
(5) and (6) is essentially
the equivalent of the FOSA.  The inclusion of higher order terms
is an additional complication which is extraneous to, and
independent of the focus on the 
back reaction in this paper:  The back reaction
reflects the specific effect of the magnetic field on the velocity, not the combination
of velocities that appear in the dynamo coefficients.

In the usual kinematic approximation, $\bfv$ is prescribed
independently of $\bfb$ and $\bfbb$.  This motivates the use of (4) to 
eliminate $\bfb$ in (3).  However, the functional forms of the dynamo coefficients are exactly valid even when $\bfv=\bfv(\bfb,\bfbb)$.  
The specific  $\bfv(\bfb,\bfbb)$ will depend on the application, but 
the field would always be inhibitive to turbulent motions.
Note that because the time scale for growth of the small scale field 
is much shorter than that of the mean field [1], the most important
back reaction comes from the small scale field.  
This is the natural interpretation of studies [7] which show effects
of the back reaction for values of the mean field much less than
equipartition with the turbulence.

{\it 3. Phases of Dynamo Growth and Flux Tube Formation-}
Observations of the sun [14,15]  and simulations of MHD turbulence in 
low $\beta_{ave}\ (\equiv {\bar P}_{part}/{\bar P}_{mag}$ where 
${\bar P_{part}}$ and  ${\bar P_{mag}}$ are the average particle and 
magnetic pressures) plasmas [16]  indicate that  the magnetic field 
tends to concentrate in flux tubes.  Intermittency in magnetic field strength in the Galactic interstellar medium is seen as well [17].  
Determining the size of flux tubes and the role of 
$\beta_{ave}$ will be addressed later but how such
intermittent structures form is addressed first.

A working 
mean field dynamo would incur phases
given an initial seed field [13].  
In the 1st phase, turbulent energy stirs up the rms random magnetic 
field to equipartition.  In the 2nd phase, in principle, 
the mean field also nears equipartition with the turbulence and/or shear 
while the small scale field remains
at or near equipartition.  In the 3rd phase, the 
dynamo works to sustain the mean field.
The back reaction is straightforwardly unimportant only during the 1st 
phase which lasts for a time $\sim \tau_{ed}$.

Flux tubes can form in phase 1 and Ref. [18] is relevant. 
There, the evolution of a seed magnetic field in homogeneous, isotropic 
turbulence is studied.  
An important result of [18] is  that 
the field tends to concentrate locally into flux tubes or ropes.
The field only grows in a local region 
when turbulence conspires to produce to proper stretching, twisting 
and folding [19].  Assuming that the mean field is constant over an eddy 
turnover time scale, $d{\bf B}/dt\sim d\bfb/dt$ so that  
the induction equation for the total field can be used to explore the growth 
of the random field for phase 1. In Lagrangian form
$$dB_i/dt=B_i \nabla_i v_j. \eqno(7)$$
As in [1], consider the initial  $B_i(0)$ to be aligned
with a line segment $\delta x_i(0)$.  Then at all later times
$\delta x_i(t)$ is aligned with $B_i(t)$.  
Eq. (8) then implies that the  length of a line segment parallel
to the field satisfies
$$dl(x,t)/dt=f(x,t)l(x,t),\eqno(8)$$
where $f(x,t)$ is a random function of the turbulent velocities.
Although $\lb l(x,t)\rb$ can be shown to increase exponentially [1], 
$l(x,t)$ is equally likely to decrease or increase at a position $x$.  
Similarly, the field would be equally likely to increase 
or decrease there, and thus a natural spatial intermittency would result.
The rms field in this picture, grows nearly to equipartition with the
volume averaged turbulent flow energy density, which
likely happens in only of order a  time $\sim \tau_{ed}$ [1] for a
modest seed field. 
The back reaction is most inhibitive after equipartition ensues, and this
is therefore the case of interest for this paper.

The growth of  field in a particular localized region
results from stretching and dragging of seed flux by a turbulent flow.
As the bundle or tube is dragged favorably for exponential growth, 
material will be inhibited from seeping into the tube from the ends since there the field is weak and  the force on particles 
opposes the direction of field line convergence [20].  
The amount of mass in a tube should thus remain the same, or decrease.  
Assuming that it remains the same, incompressibility implies a constant 
volume.
Thus the cross sectional area of the tube $\pi r_t^2\propto l^{-1}$,
and from flux freezing $B_t\propto l$.  
The edge of the tube is a current sheet, with no flow
normal to it, and thus pressure is balanced
across it :
$$\beta_{t}+1=(\beta_{ext}+1)(B_{ext}/B_t)^2,\eqno(9)$$
where $\beta_t\equiv 8\pi P_{part,t}/B_t^2$ and 
$\beta_{ext}\equiv 8\pi P_{part,ext}/B_{ext}^2$ with
$B_{ext}$ is of order the initial seed field, 
$P_{part,t}$ is the tube particle pressure, and
$P_{part,ext}$ is the external particle pressure.
Eq. (9) shows that $\beta_t << \beta_{ext}  $ since $B_t >> B_{ext}$ in
equilibrium.

Since the tube pressure is balanced 
by the external pressure, the magnetic energy density can be higher than the
turbulent energy density at the tube locations when $\be_t$ is small,
and then the volume filling fraction of the tubes would necessarily
be small.  
To see this note that the average magnetic and particle pressures satisfy 
${\bar P}_{mag}\sim f P_{mag,t}$ 
and ${\bar P}_{part}\sim (1-f)P_{part,ext}$
where $f$ is the fraction of volume occupied by magnetic flux tubes, 
$P_{mag,t}$ is the magnetic pressure in the flux tubes and
$P_{part,ext}$ is the particle pressure external to the tubes.
Thus $\beta_{ave}\sim [(1-f)/f]P_{part,ext}/P_{mag,t}=(1+\beta_t)(1-f)/f$.  
Thus $f/(1-f)=(1+\beta_t)/\beta_{ave}$.


Each energy containing (outer) scale eddy of 
wavelength $l$ stretches a tube to length $l$ and radius $r_t$.
The thickness of each tube, $r_t$, can be estimated by 
balancing  the magnetic and  turbulent eddy 
drag forces [9,21].  This gives
$$(B_t^2/4\pi r_c)(\pi r_t^2)\sim C_d\rho_{ext} v_l^2 2r_t,\eqno(10)$$
where $B_t$ is the magnitude of the field in the flux tube, 
$\rho_{ext}$ is the density outside the
flux tube, and  $C_d$ is the coefficient of turbulent drag.
Since $l$ is a wavelength, the radius of curvature $r_c$ can be estimated
by $l/4$ when the tube maximally responds to the turbulence.  In equilibrium,  
$B_t^2/8\pi\sim P_{part,ext}(1+\beta_t)$, so that $(10)$ gives, for 
$(1-f) \sim  1$,  
$$r_t/l=C_d \Ga M_l^2 / 4 \pi\sim  0.06 M_l^2(1+\be_t), \eqno(11)$$
where $\Ga$ is the adiabatic index and 
$M_l^2\equiv v_l^2/(\Ga P_{ext}/\rho_{ext})\sim \beta^{-1}_{ave}
\sim f(1+\be_t)^{-1}$, when equipartition between turbulent and magnetic
energy is assumed.  
For the last similarity in (11), $C_d$ 
was estimated from the ``drag'' crisis [21] which 
reduces $C_d < 1$ at large turbulent Reynolds number $R_l$.  
Assuming $R_l\gsim 1000$, $C_d\sim 0.4$.


{\it 4. Role of Flux Tubes-}
Once dynamo growth enters the 2nd and 3rd phases, the back reaction of 
the field on the turbulent eddy  becomes important with respect to the 
net transport of magnetic field.  In particular, the rapid growth in 
magnetic tension inhibits  it from traveling much more than $\sim l$.  
Unless reconnection with another tube can happen before the tension response, 
the tube will react back.  
Both the diffusion and helicity coefficients require inhibited 
motions of the turbulent velocity.  For example, inhibited 
turbulent diffusion would mean that an ink mark on some tube
statistically incurs zero net displacement 
from its initial location (i.e. oscillating motions)
instead of increasing its separation from the initial point with time.  

No inhibition of motions
 would mean that a tube could reconnect with a partner in the time it takes 
to move a distance $r_t$, namely a time $r_t/v_l$.  
After a reconnection, 
the tube would change one of its ends, and would then move in a 
different random direction from whence it came, before impacting another tube. 
The process would continue, enabling for example, an effective diffusion.  
The reconnection time scale is $r_t F(R_{M,t})/v_{A,t}$
where $F(R_{M,t})$ is the function of the
magnetic Reynolds number associated with the length scale $r_t$,
and $v_{A,t}$ is the tube Alfv\'en speed.
There would be no inhibition to turbulent motions when
$$v_{A,t}/v_l \sim M_l^{-1}(1+\beta_t)^{-1/2}\sim \beta_{ave}^{1/2}
/(1+\beta_t)^{1/2} > F(R_{M,t}),\eqno(12)$$
where the second similarity follows from equipartition.
For Sweet-Parker (SP) reconnection, $F(R_{M,t}) \sim  R_{M,t}^{1/2}$ 
while for Petschek (PK) reconnection $F(R_{M,t})\sim {\rm Ln}R_{M,t}$ [1]. 
Note that $R_{M,t}\sim(r_t / l)R_{M,l} = 0.06 R_{M,l}M_l^2(1+\be_t)
\sim 0.06 R_{M,l}(1+\be_t)/\beta_{ave}$ from (11), where $R_{M,l}$ is the
standard magnetic Reynolds number associated with $l$.
For the more stringent SP case, (12) then becomes 
$\beta_{ave}/(1+\be_t) > 0.25 R^{1/2}_{M,l}$ which is likely  satisfied 
in or above the solar convection zone of the sun [1].
Note that if the field were diffuse and not concentrated in flux tubes,
then then $\be_{ave}\sim \be_t$ and the inequality in (12) could 
not be satisfied.


 
Note that (11), and the line following it,  imply that 
$r_t/L \lsim f$.  But each tube fills a 
fraction $\sim r_t^2/L^2\lsim f^2$  of an eddy volume.  
Thus there are $\gsim 1/f$ tubes per eddy volume when $(1-f) \sim  1$.

{\it 5. Application to the Dynamo Coefficients-}
The simplest way to describe the effect of flux tubes, is to say that
fast reconnection significantly reduces 
the back reaction terms of the Lorentz force
on the velocity field in the equation of motion.
Thus the velocities in the dynamo coefficients would be approach their
kinematic values the more efficient the reconnection.
In the presence of equipartition magnetic flux tubes, the reduction from 
their kinematic limit would be determined by the 
fraction of tube Lorentz force that contributes to the back reaction, namely
the fraction that cannot reconnect during $\tau_{ed}$.
Equivalently, using (13), and noting that
dynamo coefficients depend on two powers of the velocity, we have
$${\tilde \a}_{br}\sim {\rm Min}\ [{\tilde \a}_{kin},\ \be_{ave}
(1+\be_t)^{-1}{\tilde \a_{kin}}/F(R_{M,t})^2], \eqno(13)$$
and
$${\tilde \beta}_{br}\sim {\rm Min}\ [{\tilde \beta}_{kin},\ \beta_{ave}
(1+\be_t)^{-1}{\tilde \beta_{kin}}/F(R_{M,t})^2], \eqno(14)$$
where the subscripts  $kin$ and $br$  refer to the kinematic values and the
values including the back reaction, respectively. 
The right sides of (13) and (14) 
are the minima of the quantities in brackets.  

It can be useful to think of the diffusion coefficient
 ${\tilde \beta}_{br}$, as measuring 
the fraction of eddy energy per mass, 
contained in motions which random walk
rather than oscillate.  For a given amount of total eddy energy,
a stronger back reaction means a higher fraction of non-zero frequency modes
[8].
To see this, note that
when the velocity is given by a stationary random field, 
$\lb\bfv(t)\cdot\bfv(t+\tau)\rb=C(\tau).$
Then the Fourier transform gives
$C(\omega)=\int^{\infty}_{-\infty}d\tau{\rm Exp}{[iwt]}\lb\bfv(t)\cdot\bfv(t+\tau)\rb.$
The zero frequency component, by comparison with (6) then satisfies
${\tilde \beta}=(1/4){\tilde C}(0).$
where ${\tilde C}(0)$ is the Fourier transform of the velocity correlation
(i.e. the power spectrum of the velocity field) evaluated at zero frequency.
The amount of eddy energy per mass 
in the zero frequency modes (i.e. the fraction contributing 
to the diffusion coefficient)
is the non-zero contribution to the diffusion coefficient

{\it 6. Discussion }
Rapid  reconnection, resulting from a concetration of
magnetic fields into low $\beta_t$ regions, may overcome the back reaction
on turbulent motions.
%
Diffusion of the mean field and helicity
 would be enabled not necessarily by removing a large amount of
field energy on the outer scale, 
but by allowing individual flux tubes to avoid recoiling back to
their points of origin.
The most important motions would be enabled on the 
energetically dominant scales. However, 
a steady forcing of the turbulence
on these outer scales would give a cascade to 
small scales as in (e.g.) [22], maintaining a constant magnetic + turbulent
energy density on the outer scale.  The irreversible dissipation  would occur
on the smallest scales.  Eqns (13) and (14) apply most effectively when 
$\beta_{ave} >>  {\rm Max}[1,\beta_t]$.  The value of $\beta_t$ determines how
effectively flux tubes evacuate and there may be a non-linear dependence
on $\beta_{ave}$.  Actual values will have to await future
simulations. 

If the shear energy were much larger than the turbulent energy and could
be dumped into the field fast enough, the magnetic energy may exceed the
turbulent energy.  Then the first similarity in
in the line following (11) would become  $\lsim$
and the inequality condition between the last terms in (12) would be
stricter than required, since the third term would be $\lsim$ the first two.

Finally, note that a similar rapid 
reconnection between evacuated tubes
 in low  $\beta_{ave}$ star forming molecular cloud regions of the ISM may 
remove material from field lines and initiate collapse [23].
In an initially uniform $\beta_{ave}<<1 $ plasma, non-linear compressional 
Alfv\'en waves clump material on the scale of the energy containing eddies
to density enhancements of order $1/\beta_{ave}$.  
The  Alfv\'en speed associated with the sparse regions
is large, allowing  rapid reconnection, closed
loops formation, and dissipation.  Thus intermittency can
also lead to fast reconnection in plasmas with $\beta_{ave} << 1$.

\noindent Thanks to G. Field for discussion.

\noindent [1] E.N. Parker, {\sl Cosmical Magnetic Fields}, (Oxford:
Clarendon Press, 1979).

\noindent [2] H.K. Moffat, {\sl Magnetic Field Generation in
Electrically Conducting Fluids}, (Cambridge:  Cambridge Univ. Press, 1978).

\noindent [3] A.A. Ruzmaikin, A.M. Shukurov, \& D.D. Sokoloff, {\sl
Magnetic Fields of Galaxies}, (Dodrecht: Klumer Press, 1988).

\noindent [4] F. Krause \& K.-H. Radler, {\sl Mean-Field Magnetohydrodynamics 
and Dynamo Theory}, (Oxford: Pergamon Press, 1980).

\noindent [5] R. Wielebinski \& F. Krause, A.\&Ap. Rev., $\bf 4$, 449 (1993).

\noindent [6] D. Elstner, R. Meinel \& R. Beck,  A.\&Ap. Supp., $\bf 94$, 
587 (1992).

\noindent [7] F. Cattaneo \& S.I. Vainshtein, Ap.J., $\bf 376$, L21
(1991); S.I. Vainshtein \& F. Cattaneo, Ap.J., $\bf 393$, 165 (1992);
Gruzinov, A.V., \& Diamond, P.H., Phys. Rev. Lett., {\bf 72} 1651 (1994);
Bhatacharjee, A.\&  Yuan, Y, Ap.J., 449, 739B, (1995).

\noindent [8] F. Cattaneo, Ap.J., $\bf 434$, 200(1994).

\noindent [9] J.H. Piddington, {\it Cosmical Magnetic Fields}, 
(Malbar:  Krieger) 1981.

\noindent [10] E.T. Vishniac, 1995, Ap. J., $\bf 446$, 724.

\noindent [11] K.-H. Radler, Astron. Nachr.,  $\bf 301$, 101 (1980).

\noindent [12] R.M. Kulsrud \& S.W. Anderson, Ap.J., $\bf 396$, 606
(1992); J.H. Piddington, {\sl Cosmic Electrodynamics}, (Malabar, FL:
Krieger Publishing, 1981).

\noindent [13] G.B. Field \& E.G. Blackman, preprint 1996.

\noindent [14] J.O. Stenflow, Sol. Phys., {\bf 32} 41, (1973).

\noindent [15] H. Zirin, {\it Astrophysics of the Sun}, (Cambridge: Cambridge University Press) 1988.

\noindent [16] A. Nordlund. et. al., Ap. J. {\bf 392} 647 (1992).

\noindent [17] G.L. Verschuur, Ap. \& Sp. Sci., $\bf 185$, 305 (1991). 

\noindent [18] A. Ruzmaikin, D. Sokoloff, \& A. Shukarov,
 M.N.R.A.S. {\bf 241}, 1 (1989).

\noindent [19] Ya. B. Zeldovich, A.A. Ruzmaikin, \& D.D. Sokoloff, {\it
Magnetic Fields in Astrophysics} (New York: Gordon \& Breach).

\noindent [20] A.O. Benz {\it Plasma Astrophysics} (Dodrecht:  Kluwer) 1993.

\noindent [21] L.D. Landau  \&  E.M. Lifshitz, 1987, $Fluid\ Mechanics$, 
(Oxford:  Pergamonn Press).



\noindent [22] P. Goldreich \& S. Sridhar, Ap.J., $\bf 438$, 763 (1995).

\noindent [23] S.H. Lubov \& J. Pringle, preprint, 1996.




\end